\documentstyle[12pt]{article}
\newcommand\be{\begin{equation}}
\newcommand\ee{\end{equation}}
\newcommand\bea{\begin{eqnarray}}
\newcommand\eea{\end{eqnarray}}
\newcommand\ket[1]{|#1\rangle}

\newcommand\braket[2]{\langle #1|#2\rangle}
\newcommand{\fatalpha}{{\bf \alpha \kern -0.44em \alpha}}
\newcommand{\fatsigma}{{\bf \sigma \kern -0.54em \sigma}}
\newcommand{\tpchi}{{\bf \chi \kern -0.35em \chi}}
\newcommand{\llambda}{{\bf \lambda \kern -0.45em \lambda}}

\renewcommand{\theequation}{\arabic{equation}}
\renewcommand{\theequation}{\thesection-\arabic{equation}}
\bibliography{plain}
\title{\bf Investigation of Continuous-Time Quantum Walk Via Modules of Bose-Mesner  and Terwilliger Algebras  }\vspace{20mm}
\author{ M. A. Jafarizadeh$^{a,b,c}$
 \thanks{E-mail:jafarizadeh@tabrizu.ac.ir}  ,
 S. Salimi$^{a,b}$
 \thanks{E-mail:shsalimi@tabrizu.ac.ir}
\\ $^a${\small Department of Theoretical Physics and Astrophysics,
Tabriz University, Tabriz 51664, Iran.} \\ $^b${\small Institute
for Studies in Theoretical Physics and Mathematics, Tehran
19395-1795, Iran.} \\ $^c${\small Research Institute for
Fundamental Sciences, Tabriz 51664, Iran.}} \pagebreak

\vspace{20mm}
\begin{document}
\maketitle \vspace{15mm}
\newpage
\begin{abstract}

 The continuous-time quantum walk on the  underlying graphs of
association schemes have been studied, via the algebraic
combinatorics structures of association schemes, namely
semi-simple  modules of their Bose-Mesner and (reference state
dependent)  Terwilliger algebras. By choosing the (walk)starting
site as a reference state, the Terwilliger algebra connected with
this choice turns the graph into the metric space with a distance
function, hence stratifies the graph  into a (d+1) disjoint union
of strata (associate classes), where the amplitudes of observing
the continuous-time quantum walk on all sites belonging to a given
stratum are the same. Using the similarity of all vertices of
underlying graph of an association scheme, it is shown that the
transition probabilities between the vertices depend only on the
distance between the vertices (kind of relations or association
classes).  Hence for a continuous-time quantum walk over a graph
associated with a given scheme with diameter $d$, we have exactly
$(d+1)$ different transition probabilities (i.e., the number of
strata or number of distinct eigenvalues of adjacency matrix).

In graphs of association schemes with known spectrum, namely with
relevant Bose-Mesner algebras of known eigenvalues and
eigenstates, the transition amplitudes and average probabilities
are given in terms of dual eigenvalues of association schemes. As
most of association schemes arise from finite groups, hence the
continuous-time walk   on generic group association schemes with
real and complex representations  have been  studied in great
details, where the transition amplitudes are given in terms of
characters of groups. Further  investigated examples are the walk
on graphs of association schemes of symmetric $S_n$, Dihedral
$D_{2m}$ and cyclic groups.

Also, following Ref.\cite{js},  the spectral distributions
connected to the highest irreducible representations of
Terwilliger algebras  of some rather important graphs, namely
distance regular graphs, have been presented. Then using spectral
distribution, the amplitudes of continuous-time quantum walk on
strongly regular graphs  such as cycle graph $C_n$ and Johnson,
and strongly regular graphs such as Petersen graphs and normal
subgroup graphs have been evaluated. Likewise, using the method
of spectral distribution, we have evaluated the amplitudes of
continuous-time quantum walk on symmetric product of trivial
association  schemes such as Hamming graphs, where their
amplitudes are proportional to the product of amplitudes of
constituent sub-graphs, and walk does not generate any
entanglement between constituent sub-graphs.

{\bf PACs Index: 03.65.Ud }
\end{abstract}
\vspace{70mm}
\newpage
\section{Introduction}
Random walks on graphs are the bases of a number of classical
algorithms. Examples include 2-SAT  (satisfiability for certain
types of Boolean formulas), graph connectivity, and finding
satisfying assignments for Boolean formulas.  It is this success
of random walks that motivated the study of their quantum analogs
in order to explore whether they might extend the set of quantum
algorithms. This has led to a number of studies. Quantum walks on
the line were examined by Nayak and Vishwanath \cite{nv}, and on
the cycle by Aharonov et al. \cite{aakv}. The latter has also
considered a number of properties of quantum walks on general
graphs\cite{js}. Two distinct types of quantum walks have been
identified:
 for the continuous-time
quantum walk a time independent Hamiltonian governs a continuous
evolution of a single particle in a Hilbert space spanned by the
vertices of a graph \cite{js,aakv,nko,fg} , while the
discrete-time quantum walk requires a quantum coin as an
additional degree of freedom in order to allow for a discrete-time
unitary evolution in the space of the nodes of a graph. The
connection between both types of quantum walks is not clear up to
now, but in both cases different topologies of the underlying
graph have been studied (see, for example,
\cite{nv,adz,dm,cfg,tbca,tbcaa}).

Different behavior of the quantum walk as compared to the
classical random walk have been reported under various
circumstances. For instance, a very promising feature of a quantum
walk on a hypercube, namely an exponentially faster hitting time
as compared to a classical random walk, has been presently found
(numerically) by Yamasaki et al. \cite{tyhk} and (analytically) by
Kempe \cite{jkem}. Indeed, first quantum algorithms based on
quantum walks which offer an (exponential) speedup over their
optimal classical counterpart have been reported in
Refs.\cite{ccdfgs,nsjkkw}.

On the other hand, the theory of association schemes has its
origin in the design of statistical experiments. The motivation
came from the investigation of special kinds of partitions of the
cartesian square of a set for the construction of partially
balanced block designs. In this context association schemes were
introduced by R. C. Bose and K. R. Nair. Although the concept of
an association scheme was introduced by Bose and Nair, the term
itself was first coined by R. C. Bose and T. Shimamoto in
\cite{rbs}. In 1973, through the work of P. Delsarte \cite{pdel}
certain association schemes were shown to play a central role in
the study of error correcting codes. This connection of
association schemes to algebraic codes, strongly regular graphs,
distance regular graphs, design theory etc., further intensified
their study. Association schemes have since then become the
fundamental, perhaps the most important objects in algebraic
combinatorics. To this regard association schemes have for some
time been studied by various people under such names as
centralizer algebras, coherent configurations, Schur rings etc.
Correspondingly, there are many different approaches to the study
of association schemes.

A further step in the study of association schemes was their
algebraization. This formulation was done by R. C. Bose and D. M.
Mesner who introduced to each association scheme a matrix algebra
generated by the adjacency matrices of the association scheme.
This matrix algebra came to be known as the adjacency algebra of
the association scheme or the Bose-Mesner algebra, after the names
of the people who introduced them. The other formulation was done
by P. Terwilliger, known as the Terwilliger algebra. This algebra
is a finite-dimensional, semisimple and is non-commutative in
general. The Terwilliger algebra has been used to study $P$- and
$Q$-polynomial schemes \cite{pt}, group schemes \cite{jmbmo,ebam},
and Doob schemes \cite{kta}.

Here in this paper, we study continuous-time quantum walk on the
underlying graphs arising  from  association schemes,  by using
their algebraic combinatorics structures,  namely  semi-simple
modules of their Bose-Mesner and (reference state dependent)
Terwilliger algebras. By choosing the (walk)starting site  as a
reference state, the Terwilliger algebra connected with this
choice turns the graph into the metric space with a  distance
function, and hence stratifies, the graph  into a (d+1) disjoint
union of strata (associate classes), where the amplitudes of
observing the continuous-time quantum walk on all sites belonging
to a given stratum are the same. Since all vertices of  underlying
graph of an association scheme are similar or they have a constant
measure of similarity, therefore the transition probabilities
between the vertices depend only on the distance between the
vertices (kind of relations or association classes). Hence for a
continuous-time quantum walk over a graph associated with a given
scheme with diameter $d$, we have exactly $(d+1)$ different
transition probabilities (i.e., the number of strata or number of
distinct eigenvalues of adjacency matrix).

In graphs of association schemes with known spectrum, namely with
relevant Bose-Mesner algebras of known eigenvalues and
eigenstates, the transition amplitudes and average probabilities
are given in terms of dual eigenvalues of association schemes. As
most of association schemes arise from finite groups, hence we
have  studied in great details continuous-time walk   on generic
group association schemes with real and complex representations,
where the transition amplitudes are given in terms of characters
of groups. Further more, as examples, we have investigated  walk
on graphs of association schemes of symmetric $S_n$, Dihedral
$D_{2m}$ and cyclic groups.

Also following Ref.\cite{js}, we have presented the spectral
distribution connected to the highest irreducible representation
of Terwilliger algebras of  some rather important graphs, namely
distance regular ones (since the Hilbert space of walk consists of
irreducible module of Terwilliger algebra with maximal dimension).
Then using the spectral distribution,  we have evaluated  the
amplitudes of continuous-time quantum walk distance regular graphs
such as cycle graph $C_n$ and Johnson, and strongly regular graphs
such as Petersen graphs and normal subgroup graphs. Likewise,
using the method of spectral distribution, we have  evaluated the
amplitudes of continuous-time quantum walk on symmetric product of
trivial association  schemes such as Hamming graphs, where their
amplitudes are proportional to the product of amplitudes of
constituent sub-graphs, and  walk does not generate any
entanglement between constituent sub-graphs.

The organization of this paper is as follows. In section 2, we
give a brief outline of association schemes,  Bose-Mesner and
Terwilliger algebras and stratification. Section 3 is devoted to
studying  continuous-time quantum walk on graphs with  known
spectrum. In section 4, we review continuous-time quantum walk on
group association scheme. In section 5, following
Ref.\cite{js,nob},  we investigate continuous-time quantum walk on
distance regular graphs via spectral distribution $\mu$ of the
adjacency matrix $A$. In section 6, we calculate the amplitudes
for continuous-time quantum walk on some graphs by using the
prescription of sections 3, 4, 5.  The paper is ended with a brief
conclusion and three appendices,  where the first appendix
consists of studying  the method of symmetrization of
non-symmetric group schemes, the second appendix contains the
proof of the lemma regarding the equality of the amplitudes
associated with the vertices belonging to the same stratum , and
the third appendix contains the list of some of the finite
distance regular graphs with their corresponding spectral
distributions, respectively.

\section{Association scheme, Bose-Mesner algebra, Terwilliger
algebra and its modules} In this section we give a brief outline
of some of the main features of association scheme, such as
adjacency matrices, Bose-Mesner algebra and Terwilliger algebra.
At the end  by choosing the (walk)starting site as a reference
state we stratify  the underlying graphs of association schemes
via the relevant Terwilliger algebra  connected with this choice.
\subsection{Association schemes}
First we recall the definition of association schemes. The reader
is referred to Ref.\cite{bailey}, for further information on
association
schemes.\\
\textbf{Definition 2.1.} (Association schemes). Let $V$ be a set
of vertices, and let $R_i(i = 0, 1,..., d)$ be nonempty relations
on $V$ (i.e., subset of $V\times V$). Let the following conditions
(1), (2), (3) and (4) be satisfied. Then the pair $Y=(V,
\{R_i\}_{0\leq i\leq d})$ consisting of a set $V$ and a set of
relations $\{R_i\}_{0\leq i\leq d}$ is called an association
scheme.\\
$(1)\;\ \{R_i\}_{0\leq i\leq d}$ is a partition of $V\times V$\\
$(2)\;\ R_0=\{(\alpha, \alpha) : \alpha\in V \}$\\
$(3)\;\ R_i=R_i^t$ for $0\leq i\leq d$, where
$R_i^t=\{(\beta,\alpha) :
(\alpha, \beta)\in R_i\} $\\
$(4)$ Given $(\alpha, \beta)\in R_k, \;\;\ p_{ij}^{k}=\mid
\{\gamma\in V : (\alpha, \beta)\in R_i \;\ and  \;\
(\gamma,\beta)\in R_j\}\mid$, where the constants $p_{ij}^{k}$ are
called the intersection numbers, depend  only on $i, j$ and $k$
and not on the choice of $(\alpha, \beta)\in R_k$.

Then the number $n$ of the vertices $V$ is called the order of the
association scheme and $R_i$ is called a relation or associate
class.

 Let
 $\Gamma=(V,R)$ denote a finite, undirected, connected graph, with
vertex set $V$, edge set $R$, path-length distance function
$\partial$, and diameter $d:=$max$\{\partial(\alpha, \beta):
\alpha, \beta\in V \}$. For all $\alpha, \beta\in V$ and all
integer $i$, we set
\begin{equation}\label{dr}
\Gamma_i=(V,R_i)=\{(\alpha, \beta): \alpha, \beta\in V:
\partial(\alpha, \beta)=i\},
\end{equation}
so that $\Gamma_i(\alpha)=\{\beta\in V: \partial(\alpha,
\beta)=i\}$. Then $\Gamma$ is distance regular graph if $\Gamma_0,
\Gamma_1, ..., \Gamma_d$ form an association scheme on $V$. In any
connected graph, if $\beta\in\Gamma_i(\alpha)$, then
\begin{equation}
\Gamma_1(\beta)\subseteq
\Gamma_{i-1}(\alpha)\cup\Gamma_i(\alpha)\cup\Gamma_{i+1}(\alpha).
\end{equation}
Hence in a distance regular graph, $p_{j1}^i=0 $ (for $i\neq 0$,
$j$ is not $\{i-1, i, i+1 \}$) and we set
\begin{equation}\label{abc}
 a_i=p_{ii}^0,\;\;\;\  b_i=p_{i-1,1}^i, \;\;\;\ c_i=p_{i+1,1}^i,
\end{equation}
 see Fig.1.
\subsection{The Bose-Mesner algebra}
Let $C$ denote the field of complex numbers. By $Mat_{V}(C)$ we
mean the $C$-algebra consisting of all matrices whose entries are
in $C$ and whose rows and columns are indexed by $V$. For each
integer $i$ ($0 \leq i\leq d$), let $A_i$ denote the matrix in
$Mat_V (C)$ with $(\alpha, \beta)$-entry
\begin{equation}
\bigl(A_{i})_{\alpha, \beta}\;=\;\cases{1 & if $\;(\alpha,
\beta)\in R_i$,\cr 0 & otherwise\cr}\qquad \qquad (\alpha, \beta
\in V).
\end{equation}
The matrices $A_i$ are called the adjacency matrices of the
association scheme. We then have $ A_0=I$(by (2) above)  and
\begin{equation}\label{li}
A_iA_j=\sum_{k=0}^{d}p_{ij}^kA_{k}
\end{equation}
(by (4) above), so $A_0, A_1, ..., A_d$ form a basis for a
commutative algebra \textsf{A} of $Mat_V(C)$, where \textsf{A} is
known as the Bose-Mesner algebra of $Y=(V, \{R_i\}_{0\leq i\leq
d})$.

Then, by using Eq.(\ref{abc}) and  Bose-Mesner algebra (\ref{li}),
for adjacency matrices of distance regular graph $\Gamma$, we have
$$
A_1A_i=c_{i-1}A_{i-1}+(a_1-b_i-c_i)A_i+b_{i+1}A_{i+1},
$$
\begin{equation}\label{dra}
A_1A_d=c_{d-1}A_{d-1}+(a_1-b_d)A_d,
\end{equation}
where $A_i$ is a polynomial in $A_1$ of degree $i$ and $A_1^i$ is
a linear combination of $I, A_1, ..., A_i$.

Since the matrices $A_i$ commute, they can be diagonalized
simultaneously (see Marcus and Minc \cite{mmmh}), that is, there
exist a matrix $S$ such that for each $A\in \textsf{A}$,
$S^{-1}AS$ is a diagonal matrix. Therefore $\textsf{A}$ is
semisimple and has a second basis $E_0,..., E_d$ (see
\cite{bailey}). These are matrices satisfying
$$
E_0 = \frac{1}{n}J
$$
$$
E_iE_j=\delta_{ij}E_i
$$
\begin{equation}\label{idem}
\sum_{i=0}^d E_i=I.
\end{equation}
The matrix $\frac{1}{n}J$ (where $J$ is the all-one matrix in
$\textsf{A}$) is a minimal idempotent (idempotent is clear, and
minimal follows from the rank($J = 1$)). The $E_i$, for ($0\leq
i,j\leq d$) are known as the primitive idempotent of $Y$. Let $P$
and $Q$ be the matrices relating our two bases for $\textsf{A}$:
$$
A_j=\sum_{i=0}^d P_{ij}E_i, \;\;\;\;\ 0\leq j\leq d,
$$
\begin{equation}\label{m2}
E_j=\frac{1}{n}\sum_{i=0}^d Q_{ij}A_i, \;\;\;\;\ 0\leq j\leq d.
\end{equation}
Then clearly
\begin{equation}\label{pq}
PQ=QP=nI.
\end{equation}
It also follows that
\begin{equation}\label{eign}
A_jE_i=P_{ij}E_i,
\end{equation}
which shows that the $P_{ij}$ (resp. $Q_{ij}$) is the $i$-th
eigenvalue(resp. the $i$-th dual eigenvalue ) of $A_j$ (resp.
$E_j$) and that the columns of $E_i$ are the corresponding
eigenvectors. Thus $m_i=$ rank($E_i$) is the multiplicity of the
eigenvalue $P_{ij}$ of $A_j$ (provided that $P_{ij}\neq P_{kj}$
for $k \neq i$). We see that $m_0=1, \sum_i m_i=n$, and
$m_i=$trace$E_i=n(E_i)_{jj}$ (indeed, $E_i$ has only eigenvalues
$0$ and $1$, so rank($E_k$) equals the sum of the eigenvalues).
Also, by \cite{jsc, ht}, the eigenvalues and dual eigenvalues
satisfy
 $$
P_{i0}=Q_{i0}=1, \;\;\;\ P_{0i}=k_i, \;\;\;\ Q_{0i}=m_i
 $$
 \begin{equation}
 m_jP_{ji}=k_iQ_{ij}, \;\;\;\;\ 0\leq i,j\leq d,
 \end{equation}
 where for all integer $i$ ($0\leq i\leq d$), set
 $k_i=p_{ii}^{0}$, and note that $k_i\neq 0$, since $R_i$ is
 non-empty.
 \subsection{The Terwilliger algebra and its modules}
We now recall the dual Bose-Mesner algebra of $Y$. Given a
\textit{base vertex}  $\alpha\in V$, for all integers $i$ define
$E^{\star}=E^{\star}(\alpha)\in Mat_V(C)$ ($0\leq i\leq d $) to be
the diagonal matrix with $(\beta, \beta)$-entry
\begin{equation}\label{e1}
\bigl(E_{i}^{\star})_{\beta, \beta}\;=\;\cases{1 & if $\;(\alpha,
\beta)\in R_i$,\cr 0 & otherwise\cr}\qquad \qquad (\alpha\in V).
\end{equation}
The matrix $E_i^\star$ is called the $i$-th dual idempotent of $Y$
with respect to $\alpha$. We shall always set $E_i^\star= 0$ for
$i < 0$ or $i> d$. From the definition, the dual idempotents
satisfy the relations
$$
\sum_{i=0}^{d}E_{i}^{\star}=I
$$
\begin{equation}
E_{i}^{\star}E_{j}^{\star}=\delta_{ij}E_{i}^{\star} \;\;\;\;\
0\leq i,j\leq d.
\end{equation}
It follows that the matrices $E_{0}^{\star},
E_{1}^{\star},...,E_{d}^{\star}$ form a basis for the subalgebra
$\textsf{A}^{\star} = \textsf{A}^{\star}(\alpha)$ of $Mat_V(R)$.
$\textsf{A}^{\star}$ is known as the dual Bose-Mesner algebra of
$Y$ with respect to $\alpha$. For each integer $i$ ($0\leq i\leq
d$), let $A_i^{\star} =A_i^{\star}(\alpha)$ denote the diagonal
matrix in $Mat_V(R)$ with $(\beta,\beta)$-entry
\begin{equation}\label{m3}
(A_i^{\star})_{(\beta,\beta)}=n(E_i)_{\alpha,\beta}  \;\;\;\;\
(\beta\in V).
\end{equation}
With reference to \cite{jsc,ht} the matrices $A_0^{\star},
A_1^{\star} ,. . . , A_d^{\star}$ form a second  basis for
$\textsf{A}^{\star}$ and satisfy
$$
A_0^{\star}=I,
$$
$$
A_i^{\star t}=A_i^{\star},
$$
$$
A_0^{\star}+A_1^{\star}+. . .+ A_d^{\star}=nE_0^{\star},
$$
\begin{equation}
 A_i^{\star}A_j^{\star}=\sum_{h=0}^{d}q_{ij}^hA_h^{\star}.
\end{equation}
Then by combining (\ref{m2}) with (\ref{e1}) and (\ref{m3}) we
have
$$
A_j^{\star}=\sum_{i=0}^{d}Q_{ij}E_i^{\star}, \;\;\;\;\ 0\leq j\leq
d,
$$
\begin{equation}
E_j^{\star}=\frac{1}{n}\sum_{i=0}^{d}P_{ij}A_i^{\star}, \;\;\;\;\
0\leq j\leq d.
\end{equation}
Let $Y =(V, {\{R_i\}}_{0\leq i\leq d})$ denote a scheme. Fix any
$\alpha\in V$ , and write $\textsf{A}^{\star}
=\textsf{A}^{\star}(\alpha)$. Let $T = T(\alpha)$ denote the
subalgebra of $Mat_V(C)$ generated by $\textsf{A}$ and
$\textsf{A}^{\star}$. We call $T$ the Terwilliger algebra of $Y$
with respect to $\alpha$.

Thus, we can define quantum decomposition for
distance regular graphs  by the following lemma:\\
\textbf{Lemma  (Terwilliger \cite{pt})}.  Let $\Gamma$ denote a
distance regular graph with diameter $d$. Fix any vertex $\alpha$
of $\Gamma$, and write $E_i^{\star} =E_i^{\star}(\alpha)$ ($0\leq
i\leq d$), $A_1=A$ and $T = T (\alpha)$. Define $A^{-} =
A^{-}(\alpha)$, $A^{0} = A^{0}(\alpha)$, $A^{+} = A^{+}(\alpha)$
by
\begin{equation}\label{dec1}
A^{-}=\sum_{i=1}^{d}E_{i-1}^{\star}AE_{i}^{\star}, \;\;\;\;\
A^{0}=\sum_{i=1}^{d}E_{i}^{\star}AE_{i}^{\star}, \;\;\;\;\ A^{+}=
\sum_{i=1}^{d}E_{i+1}^{\star}AE_{i}^{\star}.
\end{equation}
Then
\begin{equation}
A=A^{+}+A^{-}+A^{0},
\end{equation}
 where, this is the quantum decomposition of adjacency matrix $A$ such
 that,
\begin{equation}
   (A^{-})^{t} = A^{+}, \;\;\;\;\;\  (A^{0})^t =A^{0},
\end{equation}
which can be verified easily.

Let $W=C^V$ denote the vector space over $C$ consisting of column
vectors whose coordinates are indexed by $V$ and whose entries are
in $C$. We observe $Mat_V(C)$ which acts on $W$ by left
multiplication. We endow $W$ with the Hermitian inner product
$\langle  , \rangle$ which satisfies $\langle u ,v
\rangle=u^t\bar{v}$ for all $u, v \in W$ , where $t$ denotes the
transpose and - denotes the complex conjugation. For all $\beta\in
V$, let $\ket{\beta}$ denote the element of $W$ with a $1$ in the
$\beta$ coordinate and $0$ in all other coordinates. We observe
$\{\ket{\beta} | \beta\in V\}$ is an orthonormal basis for $W$.
Using (\ref{e1}) we have
\begin{equation}\label{Ter1}
E_i^{\star}W = span\{\ket{\beta}| \beta\in V, (\alpha,\beta)\in
R_i\}, \;\;\;\ 0\leq i\leq d.
\end{equation}
With the use of Eq.(\ref{Ter1})  and since $\{\ket{\beta} |
\beta\in V\}$ is an orthonormal basis for $W$, we get
\begin{equation}\label{tel}
W=E_0^{\star}W\oplus E_1^{\star}W\oplus\cdots \oplus E_d^{\star}W,
\end{equation}
where the orthogonal direct sum, with $0 \leq i \leq d$,
$E_i^{\star}$ acts on $W$ as the projection onto $E_i^{\star}W$,
similarly $W=\sum_{i=0}^{d}E_iW$ . We call $E_i^{\star}W$ the
$i$-th subconstituent of $\Gamma=(V,R)$ (distance regular graph)
with respect to $\alpha$. For $0\leq i\leq d$ we define
\begin{equation}\label{unitv2}
\ket{\phi_{i}}=\sum \ket{\beta} ,
\end{equation}
 where the sum is over all vertices $\beta\in V$ such that $(\alpha,\beta)\in
R_i$. We observe $\ket{\phi_{i}}\in E_{0}^{\star}W $.

 By a $T$-module we mean a subspace
$U\subseteq W$ such that $TU\subseteq U$. Let $U$ denote a
$T$-module. Then $U$ is said to be irreducible whenever $U$ is
nonzero and $U$ contains no $T$-modules other than $0$ and $U$.
Let $U$ denote an irreducible $T$-module. Then $U$ is the
orthogonal direct sum of the nonzero spaces among $E_{0}^{\star}U,
E_{1}^{\star}U, . . ., E_{d}^{\star}U$ (\cite{pt}, Lemma 3.4). By
the endpoint of $U$ we mean min $\{i| 0\leq i\leq d,
E_{i}^{\star}U\neq 0\}$. By the diameter of $U$ we mean $|\{i|
0\leq i\leq d, E_{i}^{\star}U\neq 0\}|- 1$. We say $U$ is thin
whenever $E_{i}^{\star}U$  has dimension at most $1$ for $0\leq
i\leq d$. There exists a unique irreducible $T$- module which has
endpoint $0$ (\cite{eeg}, Prop. 8.4). This module is called $W_0$.
For $0\leq i\leq d$ the vector $\ket{\phi_{i}}$ of
Eq.(\ref{unitv2}) is a basis for $E_{i}^{\star}W_0$ (\cite{pt},
Lemma 3.6). Therefore $W_0$ is thin with diameter $d$ such that
the module $W_0$ is orthogonal to each irreducible $T$-module
other than $W_0$ (\cite{bc}, Lem. 3.3).
\subsection{Stratification}
For $\alpha\neq \beta$ let $\partial(\alpha, \beta)$ be the
length of the shortest walk connecting $\alpha$ and $\beta$. By
definition $\partial(\alpha, \beta)=0$ for all $\alpha\in V$. The
graph becomes a metric space with the distance function
$\partial$. Note that $\partial(\alpha, \beta)=1$ if and only if
$\alpha\sim \beta$. We fix a point $o\in V$ as an origin of the
graph. Then, the graph is stratified into a disjoint union of
associate classes $\Gamma_{k}(o)$:
\begin{equation}\label{v1}
V=\bigcup_{k=0}^{d}\Gamma_{k}(o),\;\;\;\;\;\;
\Gamma_{k}(o)=\{\alpha\in V;\partial(o,\alpha)=k\}.
\end{equation}
With each associate class $\Gamma_{k}(o)$ we associate a unit
vector in $l^2(V)$ defined by
\begin{equation}\label{unitv}
\ket{\phi_{k}}=\frac{1}{\sqrt{a_k}}\sum_{\alpha\in\Gamma_{k}(o)}\ket{\alpha}\in
E_k^{\star}W ,
\end{equation}
where, $\ket{\alpha}$ denotes the eigenket of $\alpha$-th vertex
at the associate class $\Gamma_{k}(o)$ and $a_k=|\Gamma_{k}(o)|$.
 Obviously, the two sites, first and last strata,  have the same stratification.
 The closed subspace of $l^2(V)$ spanned by
$\{\ket{\phi_{k}}\}$ is denoted by $\Lambda(G)$. Since
$\{\ket{\phi_{k}}\}$ becomes a complete orthonormal basis of
$\Lambda(G)$, we often write
\begin{equation}
\Lambda(G)=\sum_{k}\oplus \textbf{C}\ket{\phi_{k}}.
\end{equation}
Let $A_i$ be the adjacency matrix of a distance regular graph
$\Gamma=(V,R)$ for vacuum state $\ket{0}$ we have
\begin{equation}\label{Foc1}
A_k\ket{0}=\sum_{\beta\in\Gamma_{k}(o)}\ket{\beta}.
\end{equation}
Also, we have
$$
A_k\ket{0}\in E_k^{\star}W
$$
\begin{equation}\label{decv1}
E_k^{\star}A_l\ket{\phi_0}=\delta_{lk}A_l\ket{\phi_0}
\end{equation}
 Then by using unit vectors $\ket{\phi_{k}}$
($\ket{0}=\ket{o}=\ket{\phi_0}$, with $o\in V$ as the fixed
origin),  and Eq.(\ref{Foc1}), (\ref{decv1})
 we have
\begin{equation}\label{Foc2}
A_k\ket{\phi_0}=\sqrt{a_k}\ket{\phi_k}.
\end{equation}

\section{Continuous-time quantum walk on graphs with known spectrum}
The continuous-time quantum walk is defined by replacing
Kolmogorov's equation with Schr\"{o}-dinger's equation.
 Continuous-time quantum walk was introduced by Farhi and
Gutmann \cite{fg} (see also \cite{cfg,mr}). Our treatment, though,
follow closely the analysis of Moore and Russell \cite{mr} which
we review next.  Let $l^2(V)$ denote the Hilbert space of
$C$-valued square-summable functions on V. With each $\alpha\in
V$ we associate a ket defined by $\ket{\alpha}$, then
$\{\ket{\alpha},\alpha\in V \}$ becomes a complete orthonormal
basis of $l^2(V)$.

For $0\leq i\leq d$ the vector $\ket{\phi_{i}}$ of
Eq.(\ref{unitv}) is a basis of $E_{i}^{\star}W_0$, where $W_0$ is
unique irreducible $T$- module which has endpoint $0$ (\cite{eeg},
Prop. 8.4).  Therefore, Hilbert space of continuous-time quantum
walk starting from a given site corresponds  to the  irreducible (
walk starting site related T-algebra) $T$- module $W_0$ with
maximal dimension. Hence other irreducible $T$- modules of
Terwilliger algebra $T$ are orthogonal  to Hilbert space of the
walk.

Let $\ket{\phi(t)}$ be a time-dependent amplitude of the quantum
process on  graph $\Gamma$. The wave evolution of the quantum
walk is
\begin{equation}
    i\hbar\frac{d}{dt}\ket{\phi(t)} = H\ket{\phi(t)},
\end{equation}
where we assume $\hbar = 1$,  and $\ket{\phi_{0}}$ is the initial
amplitude wave function of the particle. The solution is given by
$\ket{\phi_{0}(t)} = e^{-iHt} \ket{\phi_{0}}$. It is more natural
to deal with the Laplacian  of the graph, defined as $L=A-D$,
where $D$ is a diagonal matrix with entries
$D_{jj}=deg(\alpha_j)$. This is because we can view $L$ as the
generator matrix that describes an exponential distribution of
waiting times at each vertex. But on $s$-regular graphs, $D =
\frac{1}{s}I$, and since $A$ and $D$ commute, we get
\begin{equation} \label{eqn:phase-factor}
e^{-itH} = e^{-it(A-\frac{1}{d}I)} = e^{-it/d}e^{-itA}.
\end{equation}
This introduces an irrelevant phase factor in the wave evolution.
In this paper we consider $L=A=A_1$. Then using Eq.(\ref{m2}) we
have
\begin{equation}
\ket{\phi_{0}(t)}=e^{-iAt}\ket{\phi_0}=e^{-i\sum_{i=0}^{d}P_{i1}E_it}\ket{\phi_0},
\end{equation}
where  using the algebra of idempotents i.e., Eq.(\ref{idem}), the
above amplitude of wave function can be written as
\begin{equation}\label{f1}
\ket{\phi_{0}(t)}=\sum_{i=0}^{d}e^{-iP_{i1}t}E_i\ket{\phi_0}.
\end{equation}

 Now  using Eqs.(\ref{m2}),
(\ref{pq}), (\ref{unitv}) and (\ref{Foc2}), the matix elements of
idempotent operators between  eigenstates strata and eigenstates
of vertices  can be calculated as
\begin{equation}\label{DE1}
\braket{\phi_k}{E_i\mid\phi_{0}}=\braket{\phi_k}{\frac{1}{n}\sum_{l=0}^d
Q_{li}A_l\mid\phi_{0}}=\frac{1}{n}\sum_{l=0}^d
Q_{li}\braket{\phi_k}{A_l\mid\phi_{0}}=\frac{\sqrt{a_k}}{n}Q_{ki},
\end{equation}
or
\begin{equation}
\braket{\beta}{E_i\mid\phi_{0}}=\frac{1}{n}Q_{ki},
\end{equation}
for every $\ket{\beta}\in \Gamma_{k}(o)$.

Finally  multiplying (3-35) by $\ket{\beta}$ and using (3-38) we
get the following expression for the amplitude of observing the
particle at vertex $\beta$ at time $t$
\begin{equation} \label{f2}
\braket{\beta}{\phi_{0}(t)}=
\sum_{i=0}^de^{-iP_{i1}t}\braket{\beta}{
E_i\mid\phi_{o}}=\frac{1}{n}\sum_{i=0}^de^{-iP_{i1}t}Q_{ki}.
\end{equation}
Obviously the above result indicates that the amplitudes of
observing walk on vertices belonging to a given stratum are the
same. Actually one can straightforwardly deduce from formula
(3-38) that the transition probabilities between the vertices
depend only on the distance between the vertices (kind of
relations or association classes), irrespective of  which site the
walk has started. This is due to the fact that in association
schemes the coloring of underlying graphs or the set of relations
between vertices determine thoroughly everything. Hence for a
continuous-time quantum walk over a graph associated with a given
scheme with diameter $d$, we have exactly $(d+1)$ different
transition probabilities (i.e., the number of strata or number of
distinct eigenvalues of adjacency matrix).

 At the end by straightforward calculation, one can  evaluate
the average probability for  finite graphs of association schemes
as
\begin{equation}\label{f3}
\bar{P}(\beta)=\lim_{T\rightarrow
\infty}\frac{1}{T}\int_{0}^{T}P_t(\beta)
dt=\frac{1}{n^2}\sum_{i=0}^{d}Q_{ki}^2
\end{equation}
for every $\ket{\beta}\in \Gamma_{k}(o)$.

\section{Continuous-time quantum walk on group schemes} In
this section we briefly discuss continuous-time quantum walk on
group schemes with real and complex representations separately.
\subsection{Group association schemes} In order to study the
continuous-time quantum walk on group graphs, we need to  study
the group association schemes. One of the most important sources
of association schemes are groups. Let $G$ be a group acting on a
finite set $V$. Then $G$ has a natural action on $V\times V$ given
by $g(\alpha, \beta)= (g\alpha, g\beta)$ for $g\in G$ and $\alpha,
\beta \in V$. The orbits
\begin{equation}
\{(g\alpha, g\beta)|g\in G \}
\end{equation}
of $V\times V$ are called orbitals. The group $G$ is said to act
generously transitive when for every pair $(\alpha, \beta)\in
V\times V$ there is a group element $g\in G$ that exchanges
$\alpha$ and $\beta$, that is $g\alpha=\beta$ and $g\beta =
\alpha$. When $G$ acts generously transitive, the orbitals form
the relations of that association scheme. Now, in the following,
we consider the orbitals which  correspond to the conjugacy
classes of $G$.
 Let $G$ be a finite group,
$C_0=\{e\}, C_1, ..., C_d$ the conjugacy classes of $G$. Let
$G\times G$ act on $G$ with the action defined by $\beta(\alpha_1,
\alpha_2) =\alpha_1^{-1}\beta\alpha_2$ where $\beta, \alpha_1,
\alpha_2\in G$. Then the diagonal action of $G\times G$ on
$G\times G$ is given by $(\beta, \gamma)(\alpha_1, \alpha_2) =
(\beta(\alpha_1, \alpha_2), \gamma(\alpha_1,
\alpha_2))=(\alpha_1^{-1}\beta\alpha_2,
\alpha_1^{-1}\gamma\alpha_2)$. One can show that $(\beta_1,
\beta_2), (\gamma_1, \gamma_2)\in G\times G$ belong to the same
orbital of $G\times G$ if and only if $\beta_1^{-1}\beta_2,
\gamma_1^{-1}\gamma_2$ belong to the same conjugacy class of $G$.
Thus in this case the orbitals correspond to the conjugacy classes
of G. For $i=0,1,...,d$ define
\begin{equation}
R_i=\{(\alpha, \beta) | \alpha^{-1}\beta\in C_i\},
\end{equation}
then  $R_i$ are the orbitals of $G\times G$ and hence $X(G)=(G,
\{R_i\}_{0\leq i\leq d})$ becomes a commutative association scheme
and it is called the group association scheme of the finite group
$G$ \cite{ht}. We define class sum $\bar{C_i}$ for $i=0,1,...,d$
as
\begin{equation}
\bar{C_i} = \sum_{\gamma\in C_i}\gamma \in CG,
\end{equation}
then, for regular representation we have
$\bar{C_i}\ket{\alpha}=\sum_{\gamma\in C_i}\ket{\gamma\alpha}$.
Therefore in  regular representation, the classes sum $\bar{C_i}(
i=0,1,...,d  $) have the following matrix elements
\begin{equation}
\bigl(\bar{C_i})_{\alpha, \beta}\;=\;\cases{1 & if $\;(\alpha,
\beta)\in C_i$,\cr 0 & otherwise\cr}\qquad \qquad (\alpha, \beta
\in G).
\end{equation}
Comparing the above matrix elements with those of adjacency
matrices given in (2-4), we see that the classes sum are the
corresponding adjacency matrices of group association scheme with
the relation defined through conjugation. It is well known that
the classes sum of finite group $G$ form the basis of center of
its $CG$ ring which is certainly a commutative algebra, hence they
are closed under  multiplication defined in $CG$, i.e, we have
$$\bar{C_i}\bar{C_j} = \sum_{k=0}^{d} p_{ij}^{k} \bar{C_k}$$ (see
details in \cite{gjml}), where $p_{ij}^{k}\;\;( i,j,k=0,1,...,d)$
are intersection numbers of the group association scheme $X(G)$
and have the  following form:
\begin{equation}
p_{ij}^{k}=\frac{|C_i||C_j|}{|G|}
\sum_{\chi}\frac{\chi(\alpha_i)\chi(\alpha_j)\overline{\chi(\alpha_k)}}{\chi(1)},
\end{equation}
where the sum is over all the irreducible characters $\chi$ of $G$
\cite{mtom}. Therefore, the idempotents $\{E_0,E_1,...,E_d\}$ of
the group association scheme $X(G)$  are the projection operators
of $CG$-module i.e,
\begin{equation}\label{E1}
E_k=\frac{\chi_k(1)}{|G|}\sum_{\alpha\in
G}\chi_{k}(\alpha^{-1})\alpha.
\end{equation}
Thus eigenvalues of adjacency matrices of $A_k$, and idempotents
$E_k$, respectively are
\begin{equation}\label{eigen1}
P_{ik}=\frac{d_i\kappa_k}{m_i}\chi_i(\alpha_k), \;\;\;\;\
Q_{ik}=d_k\overline{\chi_k(\alpha_i)},
\end{equation}
 where $d_j=\chi_j(1)$. The above defined group scheme is in
general non-symmetric scheme and it can be symmetric provided that
we choose a group whose whole irreducible representations of
chosen group are real, such as symmetric group $S_N$. In appendix
A, we have explained how to construct a symmetric group
association scheme from a non-symmetric one.

\textbf{A.Continuous-time walk on group schemes with real representations}\\
In a finite group $G$ with real conjugacy classes $C_0=\{e\}, C_1,
..., C_d$, i.e., $C(\alpha)=C(\alpha^{-1})$ for all $\alpha\in
G$,  all irreducible  characters $\chi_i$ are  real. Thus using
(\ref{eigen1}) we can  study continuous-time quantum walk on its
underlying graph, where the amplitude of observing the particle at
stratum $k$ at time $t$, i.e, Eq.(\ref{f2}) reduces to
\begin{equation}\label{gr1}
\braket{\phi_k}{\phi_{0}(t)}=\frac{\sqrt{a_k}}{n}\sum_{i=0}^d d_i
e^{\frac{-id_i\kappa_{1}
\chi_{i}(\alpha_1)t}{m_i}}\overline{\chi_i(\alpha_k)},
\end{equation}
also, the average probabilities over large times  becomes
\begin{equation}
\bar{P}(k)=\frac{a_k}{n^2}\sum_{i=0}^{d}d_i^2|\chi_i(\alpha_k)|^{2},\;\;
k=0,1,...,d.
\end{equation}
Therefore,  the probability of observing the walk at starting
vertex, i.e., the staying probability  is
\begin{equation}
\bar{P}(0)=\frac{a_0}{n^2}\sum_{i=0}^{d}d_i^2|\chi_i(0)|^{2}
=\frac{1}{n^2}\sum_{i=0}^{d}d_i^4.
\end{equation}
As examples we will study continuous-time quantum on  $G=S_n, D_{2m}$ graphs in section 6.\\
\textbf{B.Continuous-time walk on group schemes with complex representations}\\
In  general all conjugacy classes of a given finite group are not
real,  hence  some of its irreducible representations become
complex and  consequently we encounter with directed undelying
graph or non-symmetric association scheme. But following
instruction of appendix $A$ we can generate a symmetric
association scheme out of non-symmetric association scheme. Thus
in this case, for continuous-time quantum walk, we need to use
formulas (A-v) and (A-vii) of the appendix $A$, where the
amplitude of observing the particle at stratum $k$ at time $t$,
i.e., Eq.(\ref{f2}) is
\begin{equation}
\braket{\phi_k}{\phi_0(t)} = \left\{\begin{array}{cc}
       \frac{\sqrt{a_k}}{n}\sum_{i=0}^l d_i
e^{\frac{-id_i\kappa_{1}
\chi_{i}(\alpha_1)t}{m_i}}\overline{\chi_i(\alpha_k)}  & \mbox{ for real representation } \\
            \frac{\sqrt{a_k}}{n}\sum_{i=l+1}^{\frac{d+l}{2}} d_i e^{\frac{-id_i\kappa_{1}
(\chi_{i}(\alpha_1)+\overline{\chi_{i}(\alpha_1)})t}{m_i}}
(\overline{\chi_i(\alpha_k)}+\chi_i(\alpha_k)) & \mbox{ for
non-real representation}.
            \end{array}\right.
\end{equation}
Also, the average probabilities are
\begin{equation}
\bar{P}(k) = \left\{\begin{array}{cc}
       \frac{a_k}{n^2}\sum_{i=0}^{l}d_i^2|\chi_i(\alpha_k)|^{2}
         & \mbox{ for real representation } \\
            \frac{a_k}{n^2}\sum_{i=l+1}^{\frac{d+l}{2}}d_i^2
|(\chi_i(\alpha_k)+\overline{\chi_i(\alpha_k)})|^{2}. & \mbox{ for
complex representation}.
            \end{array}\right.
\end{equation}
In this case the  staying  probability is
\begin{equation}\label{nonrel}
\bar{P}(0)=\frac{a_0}{n^2}(\sum_{i=0}^{l}d_i^2|\chi_i(0)|^{2}+4\sum_{i=l+1}^{\frac{d+l}{2}}d_i^2
|(\chi_i(0)|^{2})
=\frac{1}{n^2}(\sum_{i=0}^{l}d_i^4+4\sum_{i=l+1}^{\frac{d+l}{2}}d_i^4
).
\end{equation}
As an example we will study $G=C_n$ in section 6.

\section{Investigation of Continuous-time quantum walk on distance regular graphs via
spectral distribution  of  adjacency matrix }

Even though in general modules of Terwilliger algebra  stratifies
the underling graph of corresponding association scheme but in
case of  distance regular graphs, due to the existence of raising
and lowering operators given by (\ref{dec1}), we can have a
stratification similar to the one introduced in Ref\cite{js}. As
we will see in the following this particular kind of
stratification leads to introduction of spectral distribution for
adjacency matrix.

Here in this case, the strata states $\ket{\phi_{k}}$ are the
same as defined by (\ref{unitv}) of  subsection 2.4, but further
using  Eqs.(\ref{dra}) and (2-18) one can show that, the raising
and lowering operators given by (\ref{dec1}) act over them as
follows

\begin{equation}
A^{+}\ket{\phi_{k}}=\sqrt{\omega_{k+1}}\ket{\phi_{k+1}}, \;\;\;\
k\geq 0
\end{equation}
\begin{equation}
A^{-}\ket{\phi_{0}}=0, \;\;\
A^{-}\ket{\phi_{k}}=\sqrt{\omega_{k}}\ket{\phi_{k-1}}, \;\;\;\
k\geq 1
\end{equation}
\begin{equation}
A^{0}\ket{\phi_{k}}=(\alpha_{k+1})\ket{\phi_{k}}, \;\;\;\ k\geq 0.
\end{equation}
As mentioned in section 2,  $\ket{\phi_{k}},\; k=0,1,...,d$ form
basis for $W_0$ which is the irreducible $T$-module with maximal
dimension, therefore  all basis of  the irreducible $T$-module
$W_0$ can be obtained by repeated action of raising operator
$A^{+}$ on reference state $\ket{\phi_{0}}$  ) and we have
$\omega_k=c_{k-1}b_k$, $\alpha_{k+1}=a_1+c_k+b_k$ similar to
Ref.\cite{js}. The space $W_0$ equipped with  set of operators
$(\Gamma, A^{+},A^{-},A^{0})$ is an interacting Fock space
associated with the Jacobi sequence $\{c_{k-1}b_{k}, a_1-b_k-c_k,
\;\;\ k=1,2,...\}$.

 Usually the spectral properties of the adjacency matrix of a
graph play an important role in many branches of mathematics and
physics and the spectral distribution can be generalized in
various ways. In this work,  following Ref.\cite{nob},  the
spectral distribution $\mu$ of the adjacency matrix $A$ (where
$A=A_1$) is defined as:
\begin{equation}\label{v2}
<A^m>=\int_{R}x^{m}\mu(dx), \;\;\;\;\ m=0,1,2,...
\end{equation}
where $<.>$ is the mean value with respect to a the ground state
$\ket{\phi_0}$, and according to Ref.\cite{nob}, the $<A^m>$
coincides with the number of $m$-step walks starting and
terminating at $o$. Then the existence of a spectral distribution
satisfying (\ref{v2}) is a consequence of Hamburger's theorem,
see e.g., Shohat and Tamarkin
[\cite{st}, Theorem 1.2].\\
We may apply the canonical isomorphism from the interacting Fock
space (Hilbert space of continuous-time quantum walk starting from
a given site, i.e., strata states or more precisely $W_0$ the
irreducible $T$- module with maximal dimension starting) onto the
closed linear span of the orthogonal polynomials determined by the
Szeg\"{o}-Jacobi sequences $(\{\omega_k\},\{\alpha_k\})$, where
for distance-regular graphs, the parameters $\omega_k$ and
$\alpha_k$ are defined as
\begin{equation}
\omega_k=c_{k-1}b_{k}, \;\;\;\;\ \alpha_k=a_1-b_{k-1}-c_{k-1},
\end{equation}
where $a_k, b_k$ and $c_k$ have been introduced in the relations
(\ref{abc}).
 More precisely, the spectral distribution $\mu$ under question is
characterized by  orthogonalizing the polynomials $\{Q_n\}$
defined recursively by
$$ Q_0(x)=1, \;\;\;\;\;\
Q_1(x)=x-\alpha_1,$$
\begin{equation}\label{op}
xQ_n(x)=Q_{n+1}(x)+\alpha_{n+1}Q_n(x)+\omega_nQ_{n-1}(x),
\end{equation}
for $n\geq 1$. If such a spectral distribution is unique (for
distance-regular graphs is possible), the spectral distribution
$\mu$ is determined by the identity:
\begin{equation}\label{v3}
G_{\mu}(x)=\int_{R}\frac{\mu(dy)}{x-y}=\frac{1}{x-\alpha_1-\frac{\omega_1}{x-\alpha_2-\frac{\omega_2}
{x-\alpha_3-\frac{\omega_3}{x-\alpha_4-\cdots}}}}=\frac{Q_{n-1}^{(1)}(x)}{Q_{n}(x)}=\sum_{l=1}^{n}
\frac{B_l}{x-x_l},
\end{equation}
where $G_{\mu}(x)$ is called the Stieltjes transform and $B_l$ is
the coefficient in the Gauss quadrature formula corresponding to
the roots $x_l$ of the polynomial $Q_{n}(x)$ and where the
polynomials $\{Q_{n}^{(1)}\}$ are defined
recursively as\\
        $Q_{0}^{(1)}(x)=1$,\\
    $Q_{1}^{(1)}(x)=x-\alpha_2$,\\
    $xQ_{n}^{(1)}(x)=Q_{n+1}^{(1)}(x)+\alpha_{n+2}Q_{n}^{(1)}(x)+\omega_{n+1}Q_{n-1}^{(1)}(x)$,\\
    for $n\geq 1$.

Now if $G_{\mu}(x)$ is known, then the spectral distribution
$\mu$ can be recovered from $G_{\mu}(x)$ by means of the
Stieltjes inversion formula:
\begin{equation}\label{m1}
\mu(y)-\mu(x)=-\frac{1}{\pi}\lim_{v\longrightarrow
0^+}\int_{x}^{y}Im\{G_{\mu}(u+iv)\}du.
\end{equation}
Substituting the right hand side of (\ref{v3}) in (\ref{m1}), the
spectral distribution can be determined in terms of $x_l,
l=1,2,...$, the roots of the polynomial $Q_n(x)$, and  Guass
quadrature constant $B_l, l=1,2,... $ as
\begin{equation}\label{m}
\mu=\sum_l B_l\delta(x-x_l)
\end{equation}
(for more details see Refs.\cite{obh,st,obah,tsc}.)

Finally, using the relations(\ref{Foc2}) and orthogonal polynomial
$P_n(x)$, where they satisfy the recursion relations (\ref{dra})
and (\ref{op}), the other matrix elements $\label{cw1}
\braket{\phi_{k}}{A^m\mid \phi_0}$ can be written as
\begin{equation}\label{cw1}
\braket{\phi_{k}}{A^m\mid
\phi_0}=\frac{1}{\sqrt{a_k}}\int_{R}x^{m}P_{k}(x)\mu(dx),
\;\;\;\;\ m=0,1,2,... .
\end{equation}
Our main goal in this paper is the evaluation of amplitude for
continuous-time quantum walk by using Eq.(\ref{cw1}) such that we
have
\begin{equation} \label{v4}
\braket{\phi_{k}}{\phi_0(t)}=\frac{1}{\sqrt{a_k}}\int_{R}e^{-ixt}P_{k}(x)\mu(dx),
\end{equation}
where $\braket{\phi_{k}}{\phi_0(t)}$ is the amplitude of observing
the particle at level $k$ at time $t$. The conservation of
probability $\sum_{k=0}{\mid
\braket{\phi_{k}}{\phi_0(t)}\mid}^2=1$ follows immediately from
Eq.(\ref{v4}) by using the completeness relation of orthogonal
polynomials $P_n(x)$. Obviously evaluation of
$\braket{\phi_{k}}{\phi_0(t)}$ leads to the determination of the
amplitudes at sites belonging to the associate scheme (stratum)
$\Gamma_k(o)$. As  proved in the appendix B, the walk has the same
amplitude at all sites belonging to the same associated class (
stratum). Also, for the finite graphs, the formula (\ref{v4})
yields
\begin{equation}\label{fin}
\braket{\phi_{k}}{\phi_0(t)}=\frac{1}{\sqrt{a_k}}\sum_{l}B_le^{-ix_lt}P_{k}(x_l),
\end{equation}
where by straightforward  calculation one can  evaluate  the
average probabilities for the finite graphs which yields
\begin{equation}\label{averg1}
\bar{P}(k)=\lim_{T\rightarrow \infty}\frac{1}{T}\int_{0}^{T}\mid
\braket{\phi_{k}}{\phi_0(t)}\mid
^2dt=\frac{1}{a_{k}}\sum_{l}B_l^2P_{k}^2(x_l).
\end{equation}

\section{Examples}
Here in subsection 6.1  we study continuous-time quantum walk on
graphs with known spectrum.
\subsection{Complete graph $K_n$}
 A complete graph with $n$ vertices (denoted by $K_n$) is
a graph with $n$ vertices in which each vertex is connected to the
others (with one edge between each pair of vertices). A complete
graph is the trivial association scheme,  with the intersection
numbers
\begin{equation}
a_1=n-1, \;\;\ b_1=1, \;\;\ c_0=n-1.
\end{equation}
It is straightforward to show that its adjacency matrix $A$ has
egivenvalues $1$ and $-1$ with  degeneracies
$$
E_0=\frac{1}{n}J,
$$
\begin{equation}
E_1=\frac{1}{n}\left(\begin{array}{cccc} n-1 & -1 & \cdots & -1
\\-1 & n-1 & \cdots & -1\\ \vdots & \vdots & \vdots & \vdots \\ -1 & -1 & \cdots & n-1
 \end{array}\right),
\end{equation}
and using Eq.(\ref{f2}) we have
\begin{equation}
\braket{\beta}{\phi_{0}(t)} = \left\{\begin{array}{ll}
\frac{1}{n}(e^{-it}+(n-1)e^{\frac{it}{n-1}}) & \mbox{ for $\beta=0$ } \\
\frac{-2}{n}\sin(\frac{tn}{2(n-1)}) e^{\frac{-it(n-2)}{2(n-1)}}  &
\mbox{for $\beta\neq 0$}.
            \end{array}\right.
\end{equation}
Finally  using Eq.(\ref{f3}) we get the following expressions for
the average probabilities
\begin{equation}
\bar{P(\beta)} = \left\{\begin{array}{ll}
           1-\frac{2(n-1)}{n^2} & \mbox{ for $\beta=0$ } \\
           \frac{2}{n^2}  & \mbox{for $\beta\neq 0$}.
            \end{array}\right.
\end{equation}

 \subsubsection{Symmetric group $S_n$}
The symmetric group $S_n$ is ambivalent in the sense that
$C(\alpha)=C(\alpha^{-1})$ for all $\alpha\in S_n$, therefore its
conjugacy classes  form a symmetric association scheme (see
Fig.2).

 For group $S_n$,  conjugacy classes   are determined by the cycle
structures of elements when they are expressed in the usual cycle
notation. The useful notation for describing the cycle structure
is the cycle type $[\nu_1,\nu_2,...,\nu_n]$ , which is the listing
of number of cycles of each length (i.e, $\nu_1$ is the number of
one cycles, $\nu_2$ is that of two cycles and so on).  Thus, the
number of elements in  a  conjugacy class or stratum is given by
\begin{equation}
|C_{[\nu_1,\nu_2,...,\nu_n]}|=\frac{n!}{\nu_1!2^{\nu_2}\nu_2!...n^{\nu_n}\nu_n!}.
\end{equation}

On the other hand  a partition $\lambda$ of $n$ is a sequence
$(\lambda_1, . . . , \lambda_n)$ where $\lambda_1\geq\cdots\geq
\lambda_n$ and $\lambda_1 +\cdots+\lambda_n = n$, where in terms
of cycle types
$$
\lambda_1=\nu_1+\nu_2+\cdots+\nu_n,
$$
$$
\lambda_1=\nu_2+\nu_3+\cdots+\nu_n,
$$
$$
 \vdots
$$
\begin{equation}
\lambda_n=\nu_n.
\end{equation}

 The notation $\lambda\vdash n$  indicates that $\lambda$ is a
partition of $n$. There is one conjugacy class for each partition
$\lambda\vdash n$ in $S_n$, which consists of those permutations
having cycle structure described by $\lambda$. We denote by
$C_\lambda$ the conjugacy class of $S_n$ consisting of all
permutations having cycle structure described by $\lambda$.
Therefore the number of conjugacy classes of $S_n$, namely
diameter of its scheme is equal to the number of partitions of
$n$, which grows approximately by $
\frac{1}{4\pi\sqrt{3}}e^{\pi\sqrt{2n/3}}$.
 We
consider the case where the generating set consists of the set of
all transposition, i.e, $C_1=C_{[2,1,1,1,1...,1]}$. For the
characters at the transposition, it is known that \cite{ri}
\begin{equation}\label{eigen2}
\chi_\lambda(\alpha_1)=\frac{2!(n-2)!dim(\rho_\lambda)}{n!}\sum_{j}\left(\left(
\begin{array}{cc}
 \lambda_j \\ 2
 \end{array}
\right)- \left(
\begin{array}{cc}
 \lambda'_j \\ 2
 \end{array}
\right) \right).
\end{equation}
Here, $\lambda'$ is the partition generated by transposing the
Young diagram of $\lambda$, while $\lambda'_j$ and $\lambda_j$ are
the $j$-th components of the partitions $\lambda'$ and $\lambda$,
and $\rho_\lambda$ is the irreducible representation corresponding
to partition $\lambda$.

Then the eigenvalues of its adjacency matrix can be written as
\begin{equation}\label{eigen2}
P_{\lambda
1}=\frac{d_{\lambda}k_1}{m_\lambda}\chi_\lambda(\alpha_1)=\sum_{j}\left(\left(
\begin{array}{cc}
 \lambda_j \\ 2
 \end{array}
\right)- \left(
\begin{array}{cc}
 \lambda'_j \\ 2
 \end{array}
\right) \right).
\end{equation}
Therefore by using Eqs.(\ref{gr1}) and (\ref{eigen2}), one can
obtain the  amplitudes on symmetric groups. As an example we
obtain amplitude for associate class of conjugacy class of
$n$-cycles, as
\begin{equation}
\braket{\phi_{n}}{\phi_0(t)}=\frac{(2i\sin(nt/2))^{n-1}}{\sqrt{n
n!}},
\end{equation}
where the results thus obtained are in agreement with those of
Ref\cite{gw}.

 In the above calculation, we have used the following results for the
characters of the $n$-cycles
\[
\chi_\lambda \left((n)\right) = \left\{
\begin{array}{ll}
(-1)^{n-k} & \mbox{for $\lambda = (k,1,\ldots,1)$, $k \in \{1,\ldots,n\} $ }\\
\hspace{3mm} 0 & \mbox{otherwise}
\end{array}
\right.
\]
and
\[
\chi_{(k,1,\cdots,1)}\left(\mathrm{id} \right) =
{dim}(\rho_{(k,1,\cdots,1)})=\left(
\begin{array}{cc}
 n-1 \\ k-1
 \end{array}
\right), \;\;\;\ P_{\lambda1}=\frac{1}{2}(2nk-n^2-n).
\]
\subsubsection{Dihedral group $D_{2m}$} The dihedral group
$G=D_{2m}$ is semi-direct product of cyclic groups $Z_m$ and
$Z_2$ with corresponding generators $a$ and $b$. Hence it is
generated by generators $a$ and $b$ with following relations:
\begin{equation}
D_{2m}=\langle a, b: a^m=b^2=1, b^{-1}ab=a^{-1}\rangle.
\end{equation}

In finding its conjugacy classes, it is convenient to consider
whether $m$ is odd or even. Hence, we will study  continuous-time
quantum walk on dihedral group for  odd and even $m$, separately.

\textbf{1. m=odd} the dihedral group $D_{2m}$ has precisely
$\frac{1}{2}(m+1)$ conjugacy classes:
$$
C_0=\{1\}, C_1=\{b,ab,a^2b,...,a^{m-1}b\},
$$
\begin{equation}
C_2=\{a,a^{-1}\},...,
C_{\frac{m+1}{2}}=\{a^{(m-1)/2},a^{-(m-1)/2}\}
\end{equation}
( for more details see Ref.\cite{gjml}).

By using Eqs.(\ref{eigen1}) and (\ref{f2}) we obtain the following
amplitudes
\begin{equation}
\braket{\phi_k}{\phi_0(t)}= \left\{\begin{array}{ccc}
           \frac{1}{m}((m-1)+\cos(mt)) & \mbox{ for $k=0$ } \\
           \frac{1}{\sqrt{m}}(-i\sin(mt)) & \mbox{ for $k=1$ }\\
           \frac{\sqrt{2}}{m}(\cos(mt)-1)
             & \mbox{for
           $k=2,3,...,(m+3)/2$}.
            \end{array}\right.
\end{equation}
Uusing Eq.(\ref{f3}) we obtain the following  average transition
probabilities  to strata $k$ at time $t$
\begin{equation}
\bar{P}(k) = \left\{\begin{array}{ccc}
           \frac{1}{m^2}((m-1)^2+\frac{1}{2}) & \mbox{ for $k=0$ } \\
            \frac{1}{2m} & \mbox{ for $k=1$ } \\
             \frac{3}{m^2}& \mbox{for $k=2,3,...,(m+3)/2$},
            \end{array}\right.
\end{equation}\\
\textbf{1. m=even} the dihedral group $D_{2m}$ ($m=2l$) has
precisely $l+3$ conjugacy classes:
$$
C_0=\{1\}, C_1=\{a^l\}, C_2=\{a,a^{-1}\},...,
C_{l}=\{a^{l-1,a^{-l+1}}\}
$$
\begin{equation}
C_{l+1}=\{a^{2j}b: 0\leq j\leq l-1\}, C_{l+2}=\{a^{2j+1}b: 0\leq
j\leq l-1\}
\end{equation}
( for more details see Ref.\cite{gjml}).
 Since we will restrict our attention to $C_1$ generate group, therefore we consider
conjugacy classes as
\begin{equation}
\tilde{C_0}=C_0, \;\;\ \tilde{C_1}=C_{l+1}\cup C_{l+2}, \;\;\;\;\
\tilde{C_2}=C_1, \tilde{C_3}=C_2, \tilde{C_4}=C_3, \cdots,
\tilde{C}_{l+1}=C_{l}.
\end{equation}
In this case,   calculation of amplitudes and probabilities   is
similar to that of dihedral groups with  odd $m$.

\subsubsection{Cycle graph $C_n$}
 A cycle graph or cycle is a graph that consists of some
number of vertices connected in a closed chain. The cycle graph
with $n$ vertices is denoted by $ C_n$, where its graphical
representation cyclic group $ Z_n=\langle \alpha\rangle$, whit
$\alpha^{n}=1$. In this case, we consider the orbitals to
correspond to the conjugacy classes of cyclic group. Also, we give
$\tilde{C}=C(\alpha)\cup C(\alpha^{-1})$ for all $\alpha\in Z_n$,
therefore the relations $R_i$ form a symmetric association scheme
with $d$ classes on $C_n$ (called the conjugacy scheme of $C_n$).
Let $\omega_j=e^{2\pi ij/n}$ for $j=0,1,...,n-1$. Using the
properties of characters of cyclic group, and  Eq.(A-iv) and
(A-v), we obtain
$$
\tilde{P}_{j1}=\chi_{j}(1)+\chi_{j}(n-1)=\omega_{j}+\omega_{j}^{n-1}
=2\cos(2\pi j/n).
$$
\begin{equation}\label{E1}
\tilde{Q}_{kj}=\overline{\chi_{k}(j)}+\overline{\chi_{k}(n-j)}=2\cos(2\pi
jk/n).
\end{equation}
Therefore by using  Eq.(\ref{f2}), for strata $k$ we have
\begin{equation}
\braket{\phi_k}{\phi_0(t)}=\left\{\begin{array}{cc}
       \frac{1}{n}(e^{-it}+2\sum_{j=0}^{d}e^{-it\cos(2\pi
j/n)})  & \mbox{ for $k=0$} \\
          \frac{\sqrt{2}}{n}(e^{-it}+2\sum_{j=0}^{d}e^{-it\cos(2\pi
j/n)}\cos(2\pi jk/n))
             & \mbox{for $k=1,2,...,d$},
            \end{array}\right.
\end{equation}
where the results thus obtained are in agreement with those of
Ref.\cite{js, abtw}. Thus, one can evaluate the average
probability of  staying  at origin for large time as follows
Eq.(\ref{nonrel})
\begin{equation}
\bar{P}(0)=\frac{1}{n^2}(1+4\sum_{j=1}^{d}\cos^2(0)))
=\frac{1}{n^2}(1+4d),
\end{equation}
and using Eq.(\ref{f3}) we get the following probabilities of
transition to stratum $k$
\begin{equation}
\bar{P}(k)=\frac{2}{n^2}(1+4\sum_{j=1}^{d}\cos^2(2\pi jk/n))).
\end{equation}
Where here we have considered  odd, $n$,  and  calculation for
even $n$  is similar to that of cycle graph with the odd one.

In the remaining the part of paper we will consider the examples
of continuous-time quantum walk which can be  investigated via
spectral distributions  of their corresponding  adjacency
matrices.
\subsection{Strongly regular graphs}

A graph (simple, undirected and loopless) of order $n$ is strongly
regular with parameters $n, \kappa, \lambda, \eta$ whenever it is
not complete or edgeless and \\(i) each vertex is adjacent to
$\kappa$
vertices,\\
 (ii) for each pair of adjacent vertices there are $\lambda$
vertices adjacent to both,\\ (iii) for each pair of non-adjacent
vertices there are $\eta$ vertices adjacent to both.\\
For strongly regular graph, the intersection numbers are given by
\begin{equation}\label{stron0}
a_1=\kappa; \;\;\   b_1=1, \;\;\ b_2=\eta; \;\;\ c_0=\kappa,
\;\;\ c_1=\kappa-\lambda-1.
\end{equation}
By using  formula (\ref{m}) one can straightforwardly get the
following spectral distribution
\begin{equation}\label{stron1}
\mu=B_1\delta(x-x_1)+B_2\delta(x-x_2)+B_3\delta(x-x_3),
\end{equation}
where, we obtain  $x_i$ and $B_i$ for $i=1,2,3$ respectively as
$$
x_1=\kappa,
$$
$$
x_2=\frac{1}{2}(\lambda-\eta+\sqrt{(\lambda-\eta)^2-4(\eta-\kappa)}),
$$
\begin{equation}\label{stron2}
x_3=\frac{1}{2}(\lambda-\eta-\sqrt{(\lambda-\eta)^2-4(\eta-\kappa)}),
\end{equation}
$$
B_1=\frac{\eta}{\kappa^2-\kappa(\lambda-\eta)+(\eta-\kappa)},
$$
$$
B_2=\frac{-\kappa\sqrt{(\lambda-\eta)^2-4(\eta-\kappa)}+\kappa(\lambda-\eta)+2\kappa}
{(\lambda-\eta-2\kappa)\sqrt{(\lambda-\eta)^2-4(\eta-\kappa)}+(\lambda-\eta)^2-4(\eta-\kappa)},
$$
\begin{equation}\label{stron3}
B_3=\frac{\kappa\sqrt{(\lambda-\eta)^2-4(\eta-\kappa)}+\kappa(\lambda-\eta)+2\kappa}
{(-\lambda+\eta+2\kappa)\sqrt{(\lambda-\eta)^2-4(\eta-\kappa)}+(\lambda-\eta)^2-4(\eta-\kappa)}.
\end{equation}
Again using Eq.(\ref{v4}) one can obtain the amplitudes  for
quantum walk at strata  $k$ and time $t$.

For example we study continuous-time quantum walk on the
following  graphs.
\subsubsection{Petersen
graph}  Petersen graph is strongly regular graph with parameters
$(n, \kappa, \lambda, \eta)=(10,3,0,1)$ (see Fig.3).

The intersection numbers and spectral distribution are
$$
a_1=3, \;\;\ a_2=6; \;\;\ b_1=b_2=1; \;\;\ c_0=3, \;\;\ c_1=2.
$$
\begin{equation}
\mu=\frac{1}{10}\delta(x-3)+\frac{1}{2}\delta(x-1)+\frac{2}{5}\delta(x+2).
\end{equation}
Therefore, the amplitudes  for walk at time $t$ are
$$
\braket{\phi_{0}}{\phi_0(t)}=\int_{R}e^{-ixt}\mu(dx)=\frac{1}{2}e^{-it}+\frac{2}{5}e^{2it}+\frac{1}{10}e^{-i3t}
$$
$$
\braket{\phi_{1}}{\phi_0(t)}=\frac{1}{\sqrt{3}}\int_{R}xe^{-ixt}\mu(dx)=\frac{1}{\sqrt{3}}(\frac{1}{2}e^{-it}
-\frac{4}{5}e^{2it}+\frac{3}{10}e^{-i3t})
$$
\begin{equation}
\braket{\phi_{2}}{\phi_0(t)}=\frac{1}{\sqrt{6}}\int_{R}(x^2-3)e^{-ixt}\mu(dx)=\frac{1}{\sqrt{6}}(-e^{-it}+\frac{2}{5}e^{2it}
+\frac{2}{5}e^{-i3t}).
\end{equation}
\subsubsection{Normal subgroup graphs}
 Let $G$ be a finite group, and $P=\{P_0,P_1,...,P_d \}$ be
a blueprint of it. We always assume that the sets $P_i$ are so
numbered that the identity element $e$ of $G$ belongs to $P_0$. If
$P_0=\{e\}$, then $P$ is called homogeneous. Let
$\{R_0,R_1,...,R_d\}$ be the set of relations
$R_i=\{(\alpha,\beta)\in G\otimes G|\alpha^{-1}\beta\in P_i\}$ on
$G$. Now, we define a blueprint for group $G$ which form a
strongly regular graph. If $H$ is a subgroup of $G$, we define
 the blueprints by
\begin{equation}
P_0=\{e\}, \;\;\;\ P_1=G-\{H\}, \;\;\;\ P_2=H -\{e\}.
\end{equation}
This blueprint form a strongly regular graph with parameters $(n,
\kappa, \lambda, \eta)=(|G|,|G|-|H|,|G|-2|H|,|G|-|H|)$.

As an example, we consider $G=D_{2m}$:

\textbf{1. m=odd} in this case  the  subgroup is defined as
\begin{equation}
H=\{e,a,a^{-1},...,a^{(m-1)/2},a^{-(m-1)/2}\}.
\end{equation}
Therefore the blueprints are given by
$$
P_0=\{e\}, \;\;\ P_1=\{b,ab,a^2b,...,a^{m-1}b\},
$$
\begin{equation}
P_2=\{a,a^{-1},...,a^{(m-1)/2},a^{-(m-1)/2}\},
\end{equation}
which form  a strongly regular graph with parameters $(2m,m,0,m)$.
By using Eqs.(\ref{stron0}),(\ref{stron1}), (\ref{stron2}) and
(\ref{stron3}), we get the following expressions for the
intersection numbers and spectral distribution
$$
a_1=m, \;\;\ a_2=m-1; \;\;\ b_1=1, \;\;\ b_2=m; \;\;\ c_0=m, \;\;\
c_1=m-1.
$$
\begin{equation}
\mu=\frac{1}{2m}\delta(x-m)+\frac{m-1}{m}\delta(x)+\frac{1}{2m}\delta(x+m).
\end{equation}
Therefore, the amplitudes for walk at time $t$ are
$$
\braket{\phi_{0}}{\phi_0(t)}=\frac{1}{m}((m-1)+\cos(mt))
$$
$$
\braket{\phi_{1}}{\phi_0(t)}=\frac{1}{\sqrt{m}}(-i\sin(mt))
$$
\begin{equation}
\braket{\phi_{2}}{\phi_0(t)}=\frac{\sqrt{m-1}}{m}(\cos(mt)-1).
\end{equation}
Also using Eq.(\ref{averg1}) we evaluate the  average
probabilities as
\begin{equation}
\bar{P}(k) = \left\{\begin{array}{ccc}
           \frac{1}{m^2}((m-1)^2+\frac{1}{2}) & \mbox{ for $k=e$ } \\
            \frac{1}{2m} & \mbox{ for $k=1$ } \\
           \frac{3(m-1)}{2m^2}  & \mbox{for $k=2$}.
            \end{array}\right.
\end{equation}
\textbf{2. m=even} we consider $m=2l$ and define subgroup as
\begin{equation}
H=\{e,a,a^{-1},...,a^{(l-1)},a^{-(l-1)},a^l\}.
\end{equation}
Therefore the blueprints are given by
$$
P_0=\{e\}, \;\;\ P_1=\{a^{2j}b, a^{2j+1}b, \;\;\ 0\leq j\leq
l-1\},
$$
\begin{equation}
P_2=\{a,a^{-1},...,a^{(l-1)},a^{-(l-1)},a^l\},
\end{equation}
which are strongly regular graphs with parameters $(2m,m,0,m)$. In
this case, the calculation is similar to that of dihedral groups
with $m$, odd.

\subsection{Cycle graph $C_n$}
\textbf{spectral distribution}

 A cycle graph or cycle has already been defined in the subsection 6.1. The cycle graph $C_n$ is
the distance regular graph, and quantum walk on them turns out to
be different for odd and even n, hence, below we treat them
separately.

\textbf{Odd n}. For odd $n =2d+1$, the diameter of graph
$C_{2d+1}$ is $d$ also the intersection numbers as
\begin{equation}
a_1=a_2=\cdots=a_d=2 ; \;\;\ b_1=b_2=\cdots=b_d=1; \;\;\ c_0=2,
\;\;\ c_1=c_2=\cdots=c_{d-1}=1.
\end{equation}
In this case, the adjacency matrices are given by
\begin{equation}
A_0=I_n, \;\;\;\ A_1=S^{1}+S^{-1}, \;\;\;\
A_2=S^{2}+S^{-2},\cdots, A_d=S^{d}+S^{-d},
\end{equation}
where $S$ is the shift operator $S^{2d+1}=1$. Now by using the
Bose-Mesner algebra of distance-regular graph for adjacency
matrices (\ref{dra}), we obtain
\begin{equation}\label{cy1}
A_l=2T_{l}(A_1/2), \;\;\;\;\ l\geq 1,
\end{equation}
where $T_l$ is Tchebichef polynomials of first kind. Also,
Eq.(\ref{cy1}) and Eq.(\ref{dra}) for $i=d$ we have
\begin{equation}\label{cy2}
T_{d}(A_1/2)=T_{d+1}(A_1/2).
\end{equation}
 The canonical isomorphism from the interacting Fock
space onto the closed linear span of the orthogonal polynomials
 maps states $\ket{\phi_{l}}$ into  orthogonal polynomials $P_l(x)=2T_{l}(x/2)$.

It is straightforward to show that the spectral distribution
Eq.(\ref{m}) as
\begin{equation}\label{cy3}
\mu=\frac{1}{2d+1}\delta(x-2)+\frac{2}{2d+1}\sum_{l=1}^{d}\delta(x-2\cos(\frac{2l\pi}{2d+1}))
\end{equation}
where $x_l=2\cos(\frac{2l\pi}{2d+1})$ are the roots of
Eq.(\ref{cy2}). Therefore, the amplitudes  for observing the walk
at time $t$ and $k$th associated class is
$$
\braket{\phi_{k}}{\phi_0(t)}=\frac{2}{\sqrt{a_k}}\int_{R}e^{-ixt/2}T_{k}(x/2)
\mu(dx)
$$
\begin{equation}
=\frac{\sqrt{2}}{2d+1}(e^{-it}+2\sum_{l=1}^{d}e^{-it\cos{2l\pi/(2d+1)}}\cos(2kl\pi/(2d+1))),
\end{equation}
where the results thus obtained are in agreement with those of
Refs.\cite{js,abtw}.

\textbf{Even n}. For even $n =2d$, the diameter of graph $C_{2d}$
is $d$ and  the intersection numbers are
$$
a_1=a_2=\cdots=a_{d-1}=2, \;\;\ a_d=1, \;\;\
b_1=b_2=\cdots=b_{d-1}=1, \;\;\ b_d=2,
$$
\begin{equation}
 c_0=2, \;\;\
c_1=c_2=\cdots=c_{d-1}=1.
\end{equation}
In this case, its calculation is similar to that of cycle graph
with $n$, odd.

In the limit of large $n$, the cycle graph $C_n$ is the same as
the infinite line graph, with the intersection numbers
\begin{equation}
a_1=a_2=\cdots=2; \;\;\ b_1=b_2=\cdots=1; \;\;\ c_0=2, \;\;\
c_1=c_2=\cdots=1.
\end{equation}
Then the orthogonal polynomials are $P_l(x)=2T_{l}(x/2)$ and the
spectral distribution $\mu$ is
\begin{equation}
\mu=\frac{1}{\pi}\frac{1}{\sqrt{4-x^2}},  \;\;\;\;\;\;\ -2\leq
x\leq 2,.
\end{equation}
Then one can obtain spectral distribution and amplitude for every
vertex at the time $t$, where the results thus obtained are in
agreement with those of Ref.\cite{js}.

\subsection{Johnson graph}
Let $v,d$ be a pair of positive integers such that $d\leq v$. Put
$S=\{1,2,...,v\}$ and $V=\{x\subset S: \mid x\mid=d\}$. We say
that $x,y\in V$ are adjacent if $d-\mid x \cap y\mid=1$. Thus a
graph structure is introduced in $V$, which is called a Johnson
graph and denoted by $J(v,d)$ (see Fig.4). By symmetry we may
assume that $2d\leq v$. Consider the growing family of Johnson
graphs $J(v,d)$, where $d\rightarrow\infty$ and
$\frac{2d}{v}\rightarrow p\in (0,1]$. Then the associated
orthogonal polynomials are (for more details see Ref.\cite{obh})

\textbf{A.} for $p=1$, we have  Laguerre polynomials $L_n(x)$ with
the following recurrence formula:
$$
L_0(x)=1,
$$
$$
L_1(x)=x-1,
$$
\begin{equation}
xL_n(x)=L_{n+1}(x)+(2n+1)L_n(x)+n^2L_{n-1}(x), \;\;\;\ n\geq 1.
\end{equation}
 By using the fact that the  Laguerre polynomials
are  orthogonal polynomials with
 respect to the spectral
distribution $e^{-x}dx$  and  following paper \cite{js} we obtain
the following  amplitudes

\begin{equation}
\braket{\phi_{k}}{\phi_0(t)}=\frac{(it)^k}{(1+it)^{k+1}}.
\end{equation}
\textbf{B.} for $0\leq p\leq 1$, by modifying the Meixner
polynomials $M_n(x)$, we have the recurrence formula:
$$
M_0(x)=1,
$$
$$
M_1(x)=x,
$$
\begin{equation}
xM_n(x)=M_{n+1}(x)+\frac{2n}{\sqrt{p(2-p)}}M_n(x)+n^2M_{n-1}(x),
\;\;\;\ n\geq 1.
\end{equation}
Hence by using the fact that the  Meixner polynomials are
orthogonal polynomials with  respect to the spectral distribution
$\sum_{k=0}^{\infty}\frac{2(1-p)}{2-p}(\frac{p}{2-p})^k\delta(x-{\frac{-p+2(1-p)k}{\sqrt{p(2-p)}}})$
and Eq.(\ref{v4}) one can obtain the amplitudes. As an example, we
obtain the amplitude at the origin at time $t$ as
\begin{equation}
\braket{\phi_{0}}{\phi_0(t)}=\sum_{k=0}^{\infty}\frac{2(1-p)}{2-p}(\frac{p}{2-p})^ke^{-i\frac{-p+2(1-p)k}{\sqrt{p(2-p)}}t}.
\end{equation}

\subsection{Product of association schemes}

In this section, we recall some basic facts about the  symmetric
product of trivial schemes. (see \cite{sys} for more details.)
This product is important not only as a means of constructing new
association schemes from the old ones, but also for describing the
structure of certain schemes in term of particular sub-schemes or
schemes whose structure may already be known. Then  using
Eq.(\ref{f2}) and (\ref{v4}), we can evaluate  amplitudes of
continuous-time quantum walk on new association schemes. The
symmetric product of $d$-tuples of  trivial scheme $K_n$  with
adjacency matrices of $I_n, J_n-I_n$ is association scheme with
the following adjacency matrices ( generators of its Bose-Mesner
algebra)
$$
A_0=I_n\otimes I_n\otimes ...\otimes I_n,
$$
$$
A_1=\sum_{permutation}(J_n-I_n)\otimes I_n\otimes...\otimes I_n,
$$
$$
\vdots
$$
\begin{equation}
A_i=\sum_{permutation}\underbrace{(J_n-I_n)\otimes(J_n-I_n)...\otimes(J_n-I_n)}_{i}
\otimes I_n\otimes...\otimes I_n,
\end{equation}
where $J_n$ is  $n\times n$ matrix with all matrix elements equal
to one. This scheme is the well known  Hamming scheme with
intersection number
$$
a_i=\frac{(n-1)^id(d-1)...(d-i+1)}{i!}, \;\;\;\;\ 1\leq i\leq d,
$$
$$
 b_i=i,   \;\;\;\;\   1\leq i\leq d,
 $$
\begin{equation}
c_i=(n-1)(d-i), \;\;\;\;\   0\leq i\leq d-1,
\end{equation}
where its underlying graph is the cartesian  product of $d$-tuples
of cyclic group  $Z_n$. Following Ref. \cite{js} , the amplitudes
of walk and the spectral distribution in the symmetric product of
graphs can be obtained in terms of
 sub-graphs. Finally  we obtain  the following expression for the amplitude  at origin
 and spectral distribution
$$
\mu=\sum_{l=0}^{d}\frac{(n-1)^{d-l}d!}{n^dl!(d-l)!}\delta(x-nl+d),
$$
\begin{equation}
\braket{\phi_{0}}{\phi_0(t)}=\sum_{l=0}^{d}\frac{(n-1)^{d-l}d!}{n^dl!(d-l)!}e^{-it(nl-d)},
\end{equation}
respectively.

 Also one can show that  its idempotents
$\{E_0,E_1,...,E_d\}$ are symmetric product of $d$-tuples of
corresponding idempotents of  trivial schemes $K_n$. That is, we
have
$$
E_0=\frac{J_n}{n}\otimes \frac{J_n}{n}\otimes ...\otimes
\frac{J_n}{n},
$$
$$
E_1=\sum_{permutation}(I_n-\frac{J_n}{n})\otimes
\frac{J_n}{n}\otimes...\otimes \frac{J_n}{n},
$$
$$
\vdots
$$
\begin{equation}
E_i=\sum_{permutation}\underbrace{(I_n-\frac{J_n}{n})\otimes(I_n-\frac{J_n}{n})...\otimes(I_n-\frac{J_n}{n})}_{i}
\otimes \frac{J_n}{n}\otimes...\otimes \frac{J_n}{n}.
\end{equation}
 Therefore, for the eigenvalues $P_{ij}$, and dual ones $Q_{ij}$
 we get
$$
P_{ij}=C_{i}^{d}(C_{j}^{d})^{-1}(n-1)^{i-j}K_j(i),
$$
\begin{equation}
Q_{ij}=\frac{m_j}{k_i}C_{j}^{d}(C_{i}^{d})^{-1}(n-1)^{j-i}K_i(j),
\end{equation}
where $K_k(x)$ is the Krawtchouk polynomials defined as
\begin{equation}
K_k(x)=\sum_{i=1}^{k}C_{i}^{x}C_{k-i}^{n-x}(-1)^i(d-1)^{k-i},
\end{equation}
 and $C_{k}^{l}=\frac{l!}{k!(l-k)!}$. Then by using Eq.(\ref{eign}) we
obtain the amplitude  as
\begin{equation}
\braket{\phi_{k}}{\phi_0(t)}=\frac{\sqrt{a_k}}{n^d}\sum_{j=0}
e^{-it\frac{j(d-1)!(n-1)^{j-1}}{j!(d-j)!}}Q_{kj}.
\end{equation}

\section{Conclusion}
Continuous-time quantum walk on  underlying  graphs of the
association schemes have been  studied by using the irreducible
modules of Bose-Mesner and Terwilliger algebras connected  with
them, where the irreducible modules of Terwilliger algebra and
dual eigenvalues of association schemes play an important role.
Also Continuous-time quantum walk on distance and strongly regular
graphs are investigated by using the spectral distribution
associated to irreducible representation of  associated
Terwilliger algebras. Although the powerful method of spectral
distribution associated to irreducible representation of
Terwilliger algebra seems to work for continuous-time quantum walk
on distance regular graphs, we expect that one  can further
develop it for some non-distance regular graphs of association
schemes and discrete-time quantum walk on graphs of association
schemes. These are under investigation. Also, it seems that using
this formalism one can study the continuous-time and discrete-time
quantum walk on some graphs which lack association scheme
structure, but they have the same staratification structure as
those of scheme graphs, provided that quantum walk starts from a
distinguished site (see  Ref.\cite{js}).

 \vspace{1cm}
\setcounter{section}{0}
 \setcounter{equation}{0}
 \renewcommand{\theequation}{A-\roman{equation}}
  {\Large{Appendix A}}\\
In this appendix, we study the method of symmetrization of group
schemes of non-symmetric one. If we have $\alpha\in C_i$ but
$\alpha^{-1}$ is not in $C_i$, then the association scheme is
non-symmetric. In order to construct a symmetric group scheme
from a give non-symmetric one, we need to define the following
classes sum
\begin{equation}
\tilde{\bar{C_i}} = \left\{\begin{array}{cc}
       \bar{C_i}
         & \mbox{ for $i=0,1,...,l$, } \\
          \bar{C_i}+\bar{C_i}^{-1}   & \mbox{ for
$i=l+1,...,\frac{d+1+l}{2}$},
            \end{array}\right.
\end{equation}
where, $\tilde{C_i}$ and $\tilde{\bar{C_i}}$ for $i=0,1,...,l$ are
real and $i=l+1,...,\frac{d+1+l}{2}$, are complex.

One can show that the  above defined classes sum  yield the
following relations among themselves
$$
\tilde{\bar{C_i}}\tilde{\bar{C_j}}=
 \frac{|\tilde{C_i}||\tilde{C_j}|}{|G|}(\sum_{\nu,k=0}^{l}\frac{\chi_{\nu}(\alpha_i)\chi_{\nu}(\alpha_j)\overline{\chi_{\nu}(\alpha_k)}}{\chi_{\nu}(1)}\tilde{\bar{C_k}}
 +\frac{1}{2}\sum_{\nu,r=l+1}^{\frac{d+1+l}{2}}\frac{\chi_{\nu}(\alpha_i)\chi_{\nu}(\alpha_j)(\overline{\chi_{\nu}(\alpha_r)}+\chi_{\nu}(\alpha_r))}{\chi_{\nu}(1)}\tilde{\bar{C_r}})
$$
for  $i,j=0,1,...,l$,
$$
\tilde{\bar{C_i}}\tilde{\bar{C_j}}=
 \frac{|\tilde{C_i}||\tilde{C_j}|}{2|G|}(\sum_{\nu,k=0}^{l}\frac{\chi_{\nu}(\alpha_i)
 (\chi_{\nu}(\alpha_j)+\overline{\chi_{\nu}(\alpha_j)})\overline{\chi_{\nu}(\alpha_k)}}
 {\chi_{\nu}(1)}\tilde{\bar{C_k}}+
 $$
 $$
\sum_{\nu,r=l+1}^{\frac{d+1+l}{2}}
\frac{\chi_{\nu}(\alpha_i)(\chi_{\nu}(\alpha_j)+\overline{\chi_{\nu}(\alpha_j)})
(\overline{\chi_{\nu}(\alpha_r)}+\chi_{\nu}(\alpha_r))}{\chi_{\nu}(1)}\tilde{\bar{C_r}})
$$
for  $i=0,1,...,l$ and $j=l+1,...,\frac{d+1-l}{2}$,

$$
\tilde{\bar{C_i}}\tilde{\bar{C_j}}=
 \frac{|\tilde{C_i}||\tilde{C_j}|}{2|G|}
 (\sum_{\nu,k=0}^{l}\frac{(\chi_{\nu}(\alpha_i)+\overline{\chi_{\nu}(\alpha_i)})
 (\chi_{\nu}(\alpha_i)+\overline{\chi_{\nu}(\alpha_i)})\overline{\chi_{\nu}(\alpha_k)}}{\chi_{\nu}(1)}\tilde{\bar{C_k}}+
$$
 \begin{equation}
\sum_{\nu,r=l+1}^{\frac{d+1+l}{2}}\frac{(\chi_{\nu}(\alpha_i)+\overline{\chi_{\nu}(\alpha_i)})(\chi_{\nu}(\alpha_i)+\overline{\chi_{\nu}(\alpha_i)})
(\overline{\chi_{\nu}(\alpha_r)}+\chi_{\nu}(\alpha_r))}{\chi_{\nu}(1)}\tilde{\bar{C_r}})
\end{equation}
for  $i,j=l+1,...,\frac{d+1+l}{2}$. Therefore, the corresponding
idempotents
$\{\tilde{E_0},\tilde{E_1},...,\tilde{E}_{\frac{d+1+l}{2}}\}$ of
 group association scheme are  its irreducible $CG$-modules
projection operators, i.e
\begin{equation}
\tilde{E_k} = \left\{\begin{array}{cc}
       \frac{\chi_k(1)}{|G|}\sum_{j=0}^{l}\overline{\chi_{k}(\alpha_j)}\tilde{\bar{C_j}}
          \\
          \frac{\chi_k(1)}{|G|}\sum_{j=l+1}^{\frac{d+1+l}{2}}
          (\overline{\chi_{k}(\alpha_j)}+\chi_{k}(\alpha_j))\tilde{\bar{C_j}}
            \end{array}\right.
\end{equation}
for $k=0,1,...,l$ and
\begin{equation}
\tilde{E_k} = \left\{\begin{array}{cc}
       \frac{\chi_k(1)}{|G|}\sum_{j=0}^{l}2\overline{\chi_{k}(\alpha_j)}\tilde{\bar{C_j}}
          \\
          \frac{\chi_k(1)}{|G|}\sum_{j=l+1}^{\frac{d+1+l}{2}}
          (\overline{\chi_{k}(\alpha_j)}+\chi_{k}(\alpha_j))\tilde{\bar{C_j}}
            \end{array}\right.
\end{equation}
for $k=l+1,...,\frac{d+1+l}{2}$.

Obviously the above defined  association scheme is symmetric. It
is rather straightforward to see that  its  eigenvalues
$\tilde{P_{ik}}$ and dual ones $\tilde{Q_{ik}}$ are
\begin{equation}
\tilde{P}_{ik} \longrightarrow\left\{\begin{array}{cc}
       \frac{d_i\kappa_k}{m_i}\chi_i(\alpha_k)
         & \mbox{ for $i=0,1,...,l$, } \\
          \frac{d_i\kappa_k}{m_i}(\chi_i(\alpha_k)+\overline{\chi_i(\alpha_k)})
             & \mbox{for $i=l+1,...,\frac{d+1+l}{2}$}
            \end{array}\right.
\end{equation}
for $k=0,1,...,l$ and
\begin{equation}
\tilde{P}_{ik} \longrightarrow\left\{\begin{array}{cc}
       \frac{d_i\kappa_k}{m_i}(\chi_i(\alpha_k)+\overline{\chi_i(\alpha_k)})
         & \mbox{ for $i=0,1,...,l$, } \\
          2\frac{d_i\kappa_k}{m_i}(\chi_i(\alpha_k)+\overline{\chi_i(\alpha_k)})
             & \mbox{for $i=l+1,...,\frac{d+1+l}{2}$}
            \end{array}\right.
\end{equation}
for $k=0,1,...,\frac{d+1+l}{2}$,

\begin{equation}
\tilde{Q}_{ik} \longrightarrow\left\{\begin{array}{cc}
       d_k\overline{\chi_k(\alpha_i)}
         & \mbox{ for $i=0,1,...,l$, } \\
          d_k(\overline{\chi_k(\alpha_i)}+\chi_k(\alpha_i))
             & \mbox{for $i=l+1,...,\frac{d+1+l}{2}$},
            \end{array}\right.
\end{equation}
for $k=0,1,...,l$ and
\begin{equation}
\tilde{Q}_{ik} \longrightarrow\left\{\begin{array}{cc}
       2d_k\overline{\chi_k(\alpha_i)}
         & \mbox{ for $i=0,1,...,l$, } \\
          d_k(\overline{\chi_k(\alpha_i)}+\chi_k(\alpha_i))
             & \mbox{for $i=l+1,...,\frac{d+1+l}{2}$},
            \end{array}\right.
\end{equation}
for $k=0,1,...,\frac{d+1+l}{2}$.

In fact, eigenvalues(dual eigenvalues) are  sum of real and
non-real contributions.

In section 5, using the above prescription we have studied the
continuous-time quantum walk on cyclic groups.

\vspace{1cm} \setcounter{section}{0}
 \setcounter{equation}{0}
 \renewcommand{\theequation}{B-\roman{equation}}
  {\Large{Appendix B}}\\
 In this appendix we prove the following lemma in connection
with the equality of continuous-time quantum walk amplitudes on
the vertices belonging to the same associated class.

\textbf{Lemma 1.} Let $q_{ik}(t)$ denote the amplitude of
observing the continuous-time quantum walk at vertex $i\in
\Gamma_k(\alpha)$ at time $t$. Then for a class of
distance-regular  graphs, the amplitude $q_{ik}$ is the same for
all  vertices of associated class $k$, for all $t$.\\
 \emph{Proof}.\\
As the distance-regular graphs form an association scheme,
therefore they have corresponding  Terwilliger algebra. Also, for
Terwilliger algebra $T$, there exist a unique irreducible
$T$-module which has  reference point $0$, where we called $W_0$,
such that the unit vector Eq.(\ref{unitv}) is a basis for
$E_k^{\star}W_0$, and the irreducible $T$-module $W_0$ is
orthogonal to its other irreducible $T$-modules. Actually the
basis of  irreducible $T$-module $W_0$ can be   obtained  simply
by acting  the operator $A^{+}$  on reference state
$\ket{\phi_0}$ repeatedly and  also the other  irreducible
$T$-modules can be obtained by repeaded action of the operator
$A^{+}$ over some set of orthogonal vectors which are  orthogonal
to irreducible $T$-module $W_0$ and at the same time they vanish
under the action of the operator $A^{-}$. The remaining part of
proof  is almost similar to the proof presented in Appendix A of
Ref. \cite{js}.

\vspace{1cm} \setcounter{section}{0}
 \setcounter{equation}{0}
 \renewcommand{\theequation}{C-\roman{equation}}
  {\Large{Appendix C}}\\
\textbf{\large{List of some  important  distance regular graphs}}\\
\textbf{1. Collinearity graph  gen. octagon GO(s, t)}.

$\mu=\frac{1}{(s+1)(st+1)(s^2t^2+1)}\delta(x-s(t+1))+
   \frac{st(t+1)}{4(st+1-\sqrt{2st})(s+t+\sqrt{2st})}\delta(x-s+1-\sqrt{2st})+
   \frac{st(t+1)}{2(st+1)(s+t)}\delta(x-s+1)+\frac{st(t+1)}{4(st+1-\sqrt{2st})(s+t+\sqrt{2st})}\delta(x-s+1+\sqrt{2st})
   +\frac{s^4}{(s+1)(s+t)(s^2+t^2)}\delta(x+t+1)$.\\
Intersection numbers:\\
$c_0=s(t+1), \;\;\ c_1=st, \;\;\ c_2=st, \;\;\ c_3=st,$\\
$b_1=1, \;\;\  b_2=1, \;\;\ b_3=1, \;\;\  b_4=t+1$.\\
\textbf{2. Collinearity graph  gen. dodecagon GD(s, 1)}.

   $\mu=\frac{1}{((s+1)^2-3s) ((s+1)^2-s)(s+1)^2}\delta(x-2s)+
   \frac{s-1+\sqrt{3s}}{12((s+1)^2-3s)}\delta(x-s+1-\sqrt{3s})+ \frac{s-1-\sqrt{3s}}{12((s+1)^2-3s)}\delta(x-s+1+\sqrt{3s})
   +\frac{s-1+\sqrt{s}}{4((s+1)^2-s)}\delta(x-s+1+\sqrt{s})+\frac{s-1-\sqrt{s}}{4((s+1)^2-s)}\delta(x-s+1+\sqrt{s})
   +\frac{s^5}{((s+1)^2-3s) ((s+1)^2-s)(s+1)^2}\delta(x+2)$.\\
Intersection numbers:\\
$c_0=2s, \;\;\ c_1=s, \;\;\ c_2=s, \;\;\ c_3=s, \;\;\ c_4=s, \;\;\ c_5=s,$\\
$b_1=1, \;\;\  b_2=1, \;\;\ b_3=1, \;\;\  b_4=1, \;\;\ b_5=1,
\;\;\ b_6=2$.\\
\textbf{3. $M_{22}$ graph}.

   $\mu=\frac{7}{110}\delta(x+4)+
   \frac{3}{10}\delta(x+3)+ \frac{7}{15}\delta(x-1)
   +\frac{1}{6}\delta(x-4)+\frac{1}{330}\delta(x-7)$.\\
Intersection numbers:\\
$c_0=7, \;\;\ c_1=6, \;\;\ c_2=4, \;\;\ c_3=4,$\\
$b_1=1, \;\;\  b_2=1, \;\;\ b_3=1, \;\;\  b_4=6$.\\
\textbf{4. Incidence graph  pg$(k-1, k-1, k-1)$,  $k=4,5,7,8$}.

   $\mu=\frac{1}{2k^2}(\delta(x-k)+
   \delta(x+k))+ \frac{k-1}{k^2}\delta(x)
   +\frac{k-1}{2k}(\delta(x-\sqrt{k})+\delta(x+\sqrt{k}))$.\\
Intersection numbers:\\
$c_0=k, \;\;\ c_1=k-1, \;\;\ c_2=k-1, \;\;\ c_3=1,$\\
$b_1=1, \;\;\  b_2=1, \;\;\ b_3=k-1, \;\;\  b_4=k$.\\
\textbf{5. Coset graph doubly truncated binary Golay}.

   $\mu=\frac{21}{512}\delta(x+11)+
   \frac{35}{64}\delta(x+3)+ \frac{105}{256}\delta(x-5)
   +\frac{1}{512}\delta(x-21)$.\\
Intersection numbers:\\
$c_0=21, \;\;\ c_1=20, \;\;\ c_2=16,$\\
$b_1=1, \;\;\  b_2=2, \;\;\ b_3=12$.\\
\textbf{6. Coset graph extended ternary Golay code}.

   $\mu=\frac{8}{243}\delta(x+12)+
   \frac{440}{729}\delta(x+3)+ \frac{88}{243}\delta(x-6)
   +\frac{1}{729}\delta(x-24)$.\\
Intersection numbers:\\
$c_0=24, \;\;\ c_1=22, \;\;\ c_2=20,$\\
$b_1=1, \;\;\  b_2=2, \;\;\ b_3=12$.\\
\textbf{7. Wells graph}.

   $\mu=\frac{1}{16}\delta(x+3)+
   \frac{1}{4}(\delta(x+\sqrt{5})+\delta(x-\sqrt{5})+ \delta(x-1))
   +\frac{3}{16}\delta(x-5))$.\\
Intersection numbers:\\
$c_0=5, \;\;\ c_1=4, \;\;\ c_2=1, \;\;\ c_3=1,$\\
$b_1=1, \;\;\  b_2=1, \;\;\ b_3=4, \;\;\  b_4=5$.\\
\textbf{8. 3-Cover GQ(2, 2)}.

   $\mu=\frac{1}{9}\delta(x+3)+\frac{2}{5}\delta(x+2)+
   \frac{4}{15}\delta(x-3)+ \frac{1}{5}\delta(x-1)
   +\frac{1}{45}\delta(x-6))$.\\
Intersection numbers:\\
$c_0=6, \;\;\ c_1=4, \;\;\ c_2=2, \;\;\ c_3=1,$\\
$b_1=1, \;\;\  b_2=1, \;\;\ b_3=4, \;\;\  b_4=6$.\\
\textbf{9. Double Hoffman-Singleton}.

   $\mu=\frac{7}{25}(\delta(x+2)+\delta(x-2))+
   \frac{21}{100}(\delta(x+3)+\delta(x-3))+\frac{1}{100}(\delta(x+7)+\delta(x-7))$.\\
Intersection numbers:\\
$c_0=7, \;\;\ c_1=6, \;\;\ c_2=6, \;\;\ c_3=1, \;\;\ c_4=1,$\\
$b_1=1, \;\;\  b_2=1, \;\;\ b_3=6, \;\;\  b_4=6, \;\;\ b_5=7$.\\
\textbf{10. Foster graph}.

   $\mu=\frac{1}{9}\delta(x)+\frac{1}{5}(\delta(x+1)+
   \delta(x-1))+\frac{1}{10}(\delta(x+2)+\delta(x-2))+
   \frac{1}{90}(\delta(x+3)+\delta(x-3))+\frac{2}{15}(\delta(x+\sqrt{6})+\delta(x-\sqrt{6}))$.\\
Intersection numbers:\\
$c_0=3, \;\;\ c_1=2, \;\;\ c_2=2, \;\;\ c_3=2, \;\;\ c_4=2,  \;\;\ c_5=1, \;\;\ c_6=1, \;\;\ c_7=1,$\\
$b_1=1, \;\;\  b_2=1, \;\;\ b_3=1, \;\;\  b_4=1, \;\;\ b_5=2,
\;\;\ b_6=2, \;\;\  b_7=2, \;\;\ b_8=3$.

In  all of the above examples,  $a_k$ is defined in terms of
$b_k$ and $c_k$ as
\begin{equation} a_k=\frac{c_0c_1...c_{k-1}}{b_1b_2...b_k}.
\end{equation}

\newpage
{\bf Figure Captions}

{\bf Figure-1:} Shows edges through $\alpha$ and $\beta$ in a
distance regular graph.

{\bf Figure-2:} Shows the graph of symmetric group $S_3$.

{\bf Figure-3:} The Petersen graph.

{\bf Figure-4:} The Johnson graph $J(4,2)$.
\end{document}